\documentclass[aps,pra,twocolumn,10pt,citeautoscript,longbibliography]{revtex4-1}
\usepackage{amssymb}
\usepackage{amsmath}
\usepackage[pdftex]{graphicx}
\usepackage{dcolumn}
\usepackage{bm}
\usepackage{textcomp}
\usepackage{sidecap}

\usepackage[svgnames]{xcolor}
\definecolor{trueblue}{rgb}{0.0, 0.45, 0.81}
\definecolor{crimsonglory}{rgb}{0.75, 0.0, 0.2}
\definecolor{forestgreen}{rgb}{0.13, 0.55, 0.13}

\usepackage{hyperref}
\hypersetup{colorlinks,linkcolor={forestgreen},citecolor={crimsonglory},urlcolor={crimsonglory}}

\begin{document}


\title{Theory for Strained Graphene Beyond the Cauchy-Born Rule}

\author{M. Oliva-Leyva$^1$}
\email{mauriceoliva.cu@gmail.com}
\author{Chumin Wang$^1$}
\email{chumin@unam.mx}

\affiliation{$^1$Instituto de Investigaciones en Materiales, Universidad Nacional Aut\'{o}noma de M\'{e}xico, Apartado Postal 70-360, 04510 Mexico City, Mexico.}


\begin{abstract}
The low-energy electronic properties of strained graphene are usually obtained by transforming the bond vectors according to the Cauchy-Born rule. In this work, we derive a new effective Dirac Hamiltonian by assuming a more general transformation rule for the bond vectors under uniform strain, which takes into account the strain-induced relative displacement between the two sublattices of graphene. Our analytical results show that the consideration of such relative displacement yields a qualitatively different Fermi velocity with respect to previous reports. Furthermore, from the derived Hamiltonian, we analyze effects of this relative displacement on the local density of states and the optical conductivity, as well as the implications on the scanning tunneling spectroscopy, including external magnetic field, and optical transmittance experiments of strained graphene.
\end{abstract}

\maketitle

\section{Introduction} 

When a material is subjected to deformation, the interatomic distances change, which modulates the interactions among neighbor atoms and, as a consequence, its physical properties could be substantially modified. This idea is the base of the so-called strain engineering that research how to manipulate, in a controlled manner, the physical properties of materials by means of appropriate strain patterns. With the arrival of graphene and the new families of two-dimensional crystals, the implementation of such idea has been triggered due to the high stretchability of these materials \cite{Roldan2015,Amorim2016,Naumis2017}.

\begin{figure}[tb]
\includegraphics[width=\linewidth]{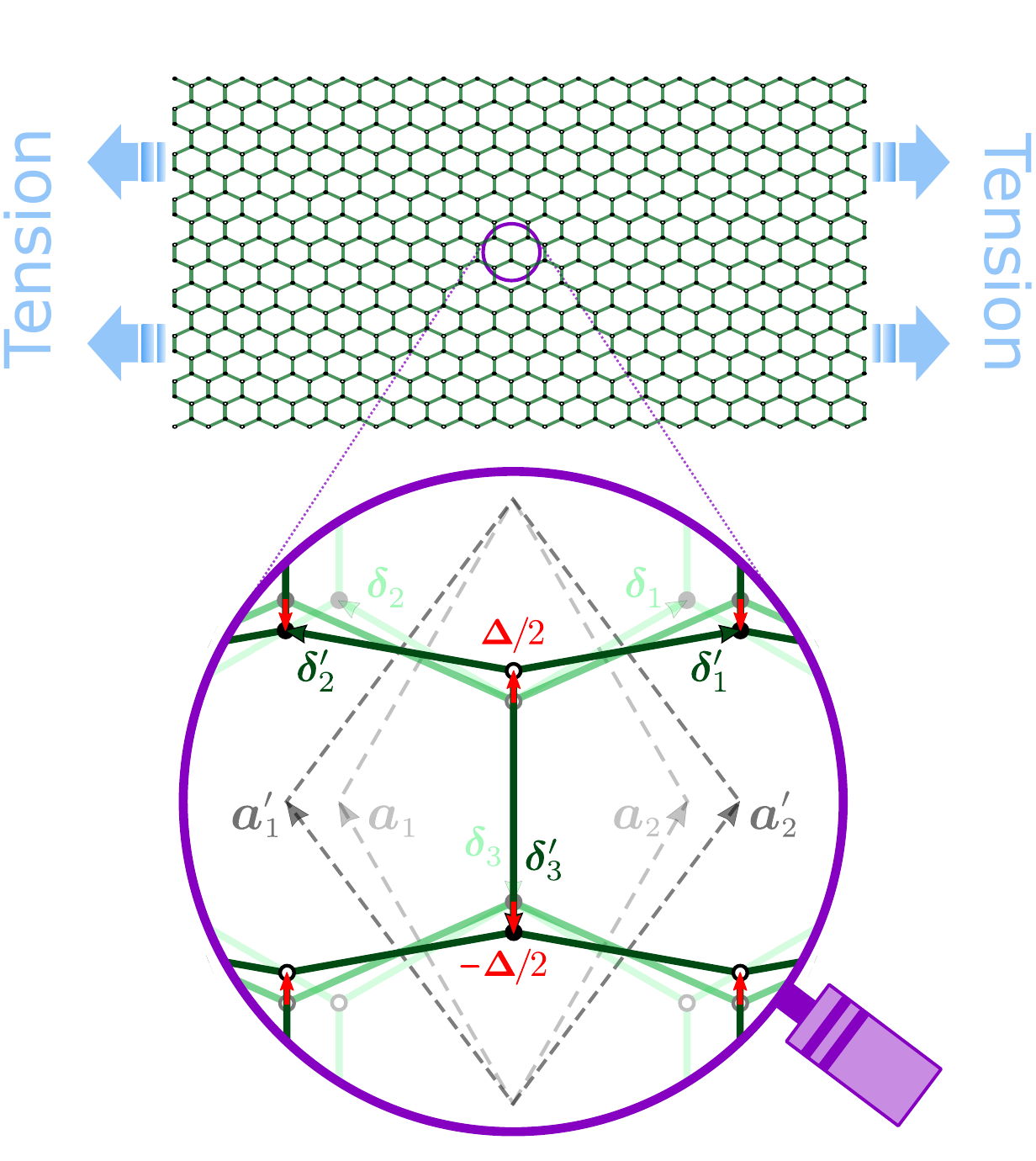}
\caption{Illustration of a portion of graphene under uniform uniaxial stretching along the zigzag direction, such that $\epsilon_{xx}>0$ and $\epsilon_{xy}=\epsilon_{yy}=0$. The zoom shows in light green color the unstrained bonds $\bm{\delta}_{n}$, in green color the strained bonds following the standard Cauchy-Born rule, and in dark green color those strained $\bm{\delta}_{n}^{\prime}$ according to equation~(\ref{GCB}) that considers the relative displacement $\bm{\Delta}$ between the two sublattices. 
The unstrained and strained unit cells are respectively defined by ($\bm{a}_{1}$, $\bm{a}_{2}$) and ($\bm{a}_{1}^{\prime}$, $\bm{a}_{2}^{\prime}$). \label{fig1}}
\end{figure}

In order to model the strain-induced effects, one needs some way of correlating macroscopic deformations (characterized by the strain tensor $\bar{\bm{\epsilon}}$) with microscopic atomic displacements. Typically in works focused on the electronic and optical properties of strained graphene \cite{Pereira2009,Pereira2010,Pellegrino2010,Pellegrino2011,Oliva2014,Oliva2013,Oliva2017a,Kaxiras2017}, this bridge is made by assuming that the undeformed nearest-neighbor vectors $\bm{\delta}_{n}$, under uniform strain, transform as the basis vectors $\bm{a}_{i}$ according to the standard Cauchy-Born rule
\begin{equation}\label{SCB}
\bm{a}_{i}^{\prime}=(\bar{\bm{I}} + \bar{\bm{\epsilon}})\cdot\bm{a}_{i},
\end{equation}
where $\bar{\bm{I}}$ is the $(2\times2)$ identity matrix. However, the deformed nearest-neighbor vectors $\bm{\delta}_{n}^{\prime}$ follow a more general rule \cite{Ericksen2008,Zhou2008,Barraza2013,Barraza2014,Midtvedt2016,Midtvedt2017}, which for graphene-like materials results as 
\begin{equation}\label{GCB}
\bm{\delta}_{n}^{\prime}=(\bar{\bm{I}} + \bar{\bm{\epsilon}})\cdot\bm{\delta}_{n} + \bm{\Delta}, 
\end{equation}
where $\bm{\Delta}$ is a relative displacement vector between the two sublattices due to additional freedom degrees introduced by the basis atoms (see Fig.~\ref{fig1}). Recently, by minimizing its strain energy parametrized in terms of the bond vectors within a valence force model, Midtvedt \emph{et al.} \cite{Midtvedt2016} obtained up to linear order in $\bar{\bm{\epsilon}}$ for graphene that 
\begin{equation}\label{d}
\bm{\Delta}= - \frac{\kappa a}{2}(2\epsilon_{xy},\epsilon_{xx}-\epsilon_{yy}),
\end{equation}
where $\kappa\approx2/5$ and $a$ is the intercarbon distance for pristine graphene. The analytical expression (\ref{d}) is referred to a Cartesian coordinate system with the $x\ (y)$ axis along the zigzag (armchair) direction of the honeycomb lattice. Note that if graphene is stretched along a direction that is perpendicular to a bond, according to the generalized Cauchy-Born rule~(\ref{GCB}) this bond changes, in contrast, it is not modified by assuming the standard Cauchy-Born rule (\ref{SCB}) for the nearest-neighbor vectors (see Fig.~\ref{fig1}). In Ref.~\cite{Midtvedt2016}, as a consequence of including $\bm{\Delta}$ in a low-energy analysis of the electronic behavior in strained graphene, it was reported that the strain-induced pseudomagnetic field keeps the same functional dependence on the strain tensor, but its strength renormalizes by a factor $(1-\kappa)\approx3/5$. 

In the presence of a uniform strain of few percent, it is important to note that the principal strain effect is to modify the Fermi velocity which becomes anisotropic \cite{Choi2010}. In fact, due to strain the Dirac cones deform from circular to elliptical cross-section. In consequence, the effective Dirac Hamiltonian for uniformly strained graphene is of the form $\mathcal{H}=\hbar \bm{\sigma}\cdot\bar{\bm{v}}\cdot\bm{q}$, where  $\bar{\bm{v}}$ is the Fermi velocity tensor, $\bm{q}$ is the momentum measured from the Dirac point and $\bm{\sigma}=(\tau\sigma_{x},\sigma_{y})$ is a Pauli matrix vector that acts on the sublattice space, with $\tau=\pm$ being the valley index. So far the previously reported expressions for $\bar{\bm{v}}$, as a function on the strain tensor, have been derived without taking into account the effect of the relative displacement vector $\bm{\Delta}$ \cite{FJ2012,FJ2013,Zubkov2014,Oliva2015a}. However, in order to gain more quantitative knowledge of the strain-induced effects on graphene, such as optical transmittance modulation \cite{Oliva2015b}, asymmetric Klein tunneling \cite{Louie2017} or dynamical gap generation \cite{Sharma2017,Xiao2017}, it is required a precise relationship between strain and the fermion velocity anisotropy. 

The main objective of this paper is to provide a low-energy Hamiltonian for strained graphene within the  generalized Cauchy-Born rule (\ref{GCB}).

\section{Effective Dirac Hamiltonian}

A standard approach to obtain the effective Dirac Hamiltonian for graphene under uniform strain is as follows. As a starting point, we use the nearest-neighbor tight-binding Hamiltonian, which can be represented in momentum space by a $(2\times2)$ matrix of the form
\begin{equation}\label{TBH}
H(\bm{k})=
\left(
\begin{array}{cc}
0 & h(\bm{k})\\
h^{\ast}(\bm{k}) & 0
\end{array}
\right),   
\end{equation}
where $h(\bm{k})=-\sum_{n=1}^{3}t_{n} e^{-i\bm{k}\cdot\bm{\delta}_{n}^{\prime}}$, the deformed nearest-neighbor vectors $\bm{\delta
}_{n}^{\prime}$ are given by equation~(\ref{GCB}) and $t_{n}$ are the modified nearest-neighbor hopping parameters. Usually the strain-induced changes of the nearest-neighbor hopping parameters are described by the exponential model $t_{n}=t e^{-\beta[(\vert\bm{\delta
}_{n}^{\prime}\vert/a)-1]}$, where $\beta\approx3$ and $t$ is the hopping parameter for pristine graphene \cite{Pereira2009,Naumis2017}. Expanding the last expression of $t_{n}$ up to linear order in the strain tensor, which is the leading order used throughout the rest of the paper, one finds  
\begin{equation}\label{tn}
t_{n}=t\hspace{.5mm}\big[1-\beta \bm{\delta}_{n}\cdot\bar{\bm{\epsilon}}\cdot\bm{\delta}_{n} - \beta \bm{\delta}_{n}\cdot\bm{\Delta}\big].
\end{equation}


Then, to obtain the effective Dirac Hamiltonian one should expand the tight-binding Hamiltonian (\ref{TBH}) around a Dirac point $\bm{K}_{D}$ \cite{Zubkov2014,Oliva2015a}. Thus, an important step within the derivation is the knowledge of the position of $\bm{K}_{D}$ which is determined by the equation, $E(\bm{K}_{D})=0$, where $E(\bm{k})=\pm\vert h(\bm{k})\vert$ is the dispersion relation resulting from Hamiltonian (\ref{TBH}). Solving $E(\bm{K}_{D})=0$, the strain-induced shift of $\bm{K}_{D}$ from the corresponding corner $\bm{K}_{0}$ of the first Brillouin zone can be expressed as
\begin{equation}\label{KD}
\bm{K}_{D}= (\bar{\bm{I}}-\bar{\bm{\epsilon}})\cdot\bm{K}_{0} + \tau\bm{A},
\end{equation}
where
\begin{eqnarray}
\bm{A}=\frac{\beta(1-\kappa)}{2a}(\epsilon_{xx}-\epsilon_{yy},-2\epsilon_{xy}),\label{A}
\end{eqnarray}
and $\tau$ is the valley index of $\bm{K}_{0}$. The expression (\ref{KD}) for $\bm{K}_{D}$ only differs from the derived one in Ref.~\cite{Oliva2013} with $\bm{\Delta}=0$ in that the vector $\bm{A}$, an emergent gauge field for nonuniform deformations \cite{Sasaki2008,Vozmediano2010}, is renormalized by a factor $(1-\kappa)$. This result confirms that previously obtained in Ref.~\cite{Midtvedt2016}. In other words, the position (\ref{KD}) of $\bm{K}_{D}$ can be obtained by replacing $\beta$ by $\beta(1-\kappa)$ in the expression of $\bm{K}_{D}$ derived without taking into account the effect of the relative displacement vector $\bm{\Delta}$ \cite{Oliva2013,Oliva2015a}.

Once the position of the Dirac point $\bm{K}_{D}$ is found, we perform the expansion of the Hamiltonian (\ref{TBH}) around $\bm{K}_{D}$, by means of $\bm{k}=\bm{K}_{D}+\bm{q}$, and we obtain that the effective Dirac Hamiltonian reads as 
\begin{equation}\label{GDH}
\mathcal{H}=\hbar \bm{\sigma}\cdot\bar{\bm{v}}\cdot\bm{q}, 
\end{equation}
where the Fermi velocity tensor $\bar{\bm{v}}$ is given by
\begin{equation}\label{TFV}
\bar{\bm{v}}=v_{0}\hspace{.5mm}\Big[\bar{\bm{I}} + \bar{\bm{\epsilon}} - \beta\bar{\bm{\epsilon}} + \beta\kappa\bar{\bm{\epsilon}} -\frac{\beta\kappa}{2}\mbox{tr}(\bar{\bm{\epsilon}})\bar{\bm{I}}\Big],
\end{equation} 
with $v_{0}=3ta/2\hbar$ being the Fermi velocity of pristine graphene. 

Let us make some important remarks about the generalized Fermi velocity tensor (\ref{TFV}). First of all, the tensorial character of $\bar{\bm{v}}$ reflects the elliptic shape of the equienergy contours around $\bm{K}_{D}$ (see Fig.\ref{fig2}~(a)). In particular, it is worth mentioning that the principal axes of $\bar{\bm{v}}$ are collinear with the ones of the strain tensor $\bar{\bm{\epsilon}}$, because the electronic anisotropy is only caused by the deformation. For example, to reveal the trigonal anisotropy due to the underlying honeycomb lattice is needed a study up to second order in the strain tensor \cite{Oliva2017a}. On the other hand, note that making $\kappa=0$ reduces equation (\ref{TFV}) to 
$\bar{\bm{v}}= v_{0}\hspace{0.5mm}[\bar{\bm{I}} + \bar{\bm{\epsilon}} - \beta\bar{\bm{\epsilon}}]$, which is the Fermi velocity tensor derived without considering $\bm{\Delta}$ \cite{Oliva2013,Oliva2015a}. At the same time, one can see that the generalized Fermi velocity tensor (\ref{TFV}) can not be obtained by making the replacement $\beta\rightarrow\beta(1-\kappa)$ in $\bar{\bm{v}}= v_{0}\hspace{0.5mm}[\bar{\bm{I}} + \bar{\bm{\epsilon}} - \beta\bar{\bm{\epsilon}}]$. In fact, the additional term $-v_{0}\beta\kappa\mbox{tr}(\bar{\bm{\epsilon}})\bar{\bm{I}}/2$ in equation (\ref{TFV}) leads to a qualitatively different behavior of the Fermi velocity as a function the strain tensor. 

To illustrate this issue, let us to consider graphene subjected a uniaxial strain of stretching magnitude $\varepsilon$ along an arbitrary direction. According to the approximation $\bar{\bm{v}}= v_{0}\hspace{0.5mm}[\bar{\bm{I}} + \bar{\bm{\epsilon}} - \beta\bar{\bm{\epsilon}}]$, the Fermi velocity perpendicular to the stretching direction is given by $v_{\perp}=v_{0}\hspace{0.5mm}[1+(\beta-1)\nu\varepsilon]$, where $\nu$ is the Poisson ratio \cite{Pereira2009,Oliva2017a}. Therefore, $v_{\perp}$ slightly increases with the increasing of $\varepsilon$. However, from the more general expression (\ref{TFV}), it follows that $v_{\perp}=v_{0}[1+(\beta-1)\nu\varepsilon - \beta\kappa(1+\nu)\varepsilon/2]$. But if $(\beta-1)\nu < \beta\kappa(1+\nu)/2$ as occurred for graphene \cite{Midtvedt2016}, then $v_{\perp}$ slightly decreases with the increasing of the stretching magnitude $\varepsilon$. Such fingerprint of the generalized Cauchy-Born rule (\ref{GCB}) on the Fermi velocity seems to be found by previous first-principles calculations of strained graphene \cite{Choi2010}, but a more detailed analysis for small strains is required.

\section{Effects on scanning tunneling spectroscopy}

From an experimental point of view, the strain-induced variations of the Fermi velocity can be measured by scanning tunneling spectroscopy (STS) \cite{Hui2013,Jang2014}, because this technique is sensitive to the local density of states (LDOS) which in turn depends on the Fermi velocity. For a strained (anisotropic) two-dimensional Dirac material described by a Hamiltonian of the form (\ref{GDH}) with a generic Fermi velocity tensor, its LDOS is given by \cite{Oliva2014,FJ2013}, 
\begin{equation}\label{LDOS}
\rho(E)=\rho_{0}(E)/\mbox{det}(\bar{\bm{v}}/v_{0}),
\end{equation}
where $\rho_{0}(E)=2\vert E\vert/(\pi\hbar^{2}v_{0}^{2})$ is the LDOS of the unstrained (isotropic) two-dimensional Dirac material with $\bar{\bm{v}}=v_{0}\bar{\bm{I}}$. Then substituting equation~(\ref{TFV}) into equation~(\ref{LDOS}) and expanding up to linear order in the strain tensor, we find that the LDOS of strained graphene reads
\begin{equation}\label{LDOS2}
\rho(E)=\rho_{0}(E)\Big[1+(\beta-1)\mbox{tr}(\bar{\bm{\epsilon}})\Big],
\end{equation}
which does not depend on $\kappa$ and exactly coincides with that obtained in Refs.~\cite{Oliva2014,FJ2013}. Therefore, $\bm{\Delta}$ does not affect the LDOS as illustrated in Fig. \ref{fig2}~(b), so that STS measurements are insensitive to the strain-induced relative displacement $\bm{\Delta}$ between the two sublattices, which is  somewhat unexpected given the additional change in the hopping parameters. 

Otherwise, STS experiments of graphene in the presence of an external magn­etic field can also be used to search the strain-induced variations of the Fermi velocity \cite{Andrei2012,Yin2017,Oliva2018a}. The most remarkable feature of these STS spectra is a series of well defined peaks at the Landau level energies, whose strain-induced shifts can be correlated with the Fermi velocity variations \cite{Andrei2012,Yin2017,Oliva2018a}. In general, for a generic anisotropic Dirac material (\ref{GDH}) in an external magn­etic field $B$, its Landau levels are given by \cite{Oliva2017b},
\begin{equation}\label{LL}
E_{n}=E_{n}^{(0)}\sqrt{\mbox{det}(\bar{\bm{v}}/v_{0})},
\end{equation}
where $E_{n}^{(0)}$ correspond to those of an isotropic Dirac material. 
Replacing $\bar{\bm{v}}$ into equation~(\ref{LL}) according to the expression~(\ref{TFV}), we arrive that the Landau level spectrum of strained graphene is 
\begin{equation}\label{LL2}
E_{n}=E_{n}^{(0)}\Big[1-(\beta-1)\mbox{tr}(\bar{\bm{\epsilon}})/2\Big],
\end{equation}
which does not show dependence on $\kappa$. As a consequence, $\bm{\Delta}$ does not produce any additional shift of the LDOS peaks of strained graphene under magnetic field. Hence, Landau level spectroscopy also does not record observable effects of $\bm{\Delta}$, at least up to linear order in the strain.   

These findings seen irrelevant, however, they suggest that the scanning tunneling spectroscopy could be an appropriate technique to experimentally determine the parameter $\beta$, with total independence of the parameter $\kappa$. As discussed below, the knowledge of $\beta$ is a prerequisite to probe the effects of $\bm{\Delta}$ from transmittance experiments of strained graphene.

\section{Effects on optical measurements}

Let us further explore the effect of $\bm{\Delta}$ on the optical properties of strained graphene. As documented in Ref.~\cite{Oliva2017b}, the optical response of an anisotropic Dirac material (\ref{GDH}) can be expressed by the conductivity tensor 
\begin{equation}\label{OCT}
\bar{\bm{\sigma}}(\omega) = \sigma_{0}(\omega)\Bigg[\frac{\mbox{tr}(\bar{\bm{v}})}{\mbox{det}(\bar{\bm{v}})}\bar{\bm{v}} -\bar{\bm{I}} \Bigg],
\end{equation}
where $\sigma_{0}(\omega)$ is the frequency-dependent optical conductivity of the isotropic Dirac material with $\bar{\bm{v}}=v_{0}\bar{\bm{I}}$. Once again, making the substitution into equation~(\ref{OCT}) of $\bar{\bm{v}}$ by expression~(\ref{TFV}), the optical conductivity tensor of strained graphene up to first order in the strain tensor $\bar{\bm{\epsilon}}$ results
\begin{equation}\label{OCT-G}
\bar{\bm{\sigma}}(\omega)=\sigma_{0}(\omega) \Big[\bar{\bm{I}} - 2\beta^{*}\bar{\bm{\epsilon}} + \beta^{*}\mbox{tr}(\bar{\bm{\epsilon}})\bar{\bm{I}}\Big],
\end{equation}
where $\beta^{*}=\beta(1-\kappa)-1$. 

\begin{figure*}[tb]
\includegraphics[width=\linewidth]{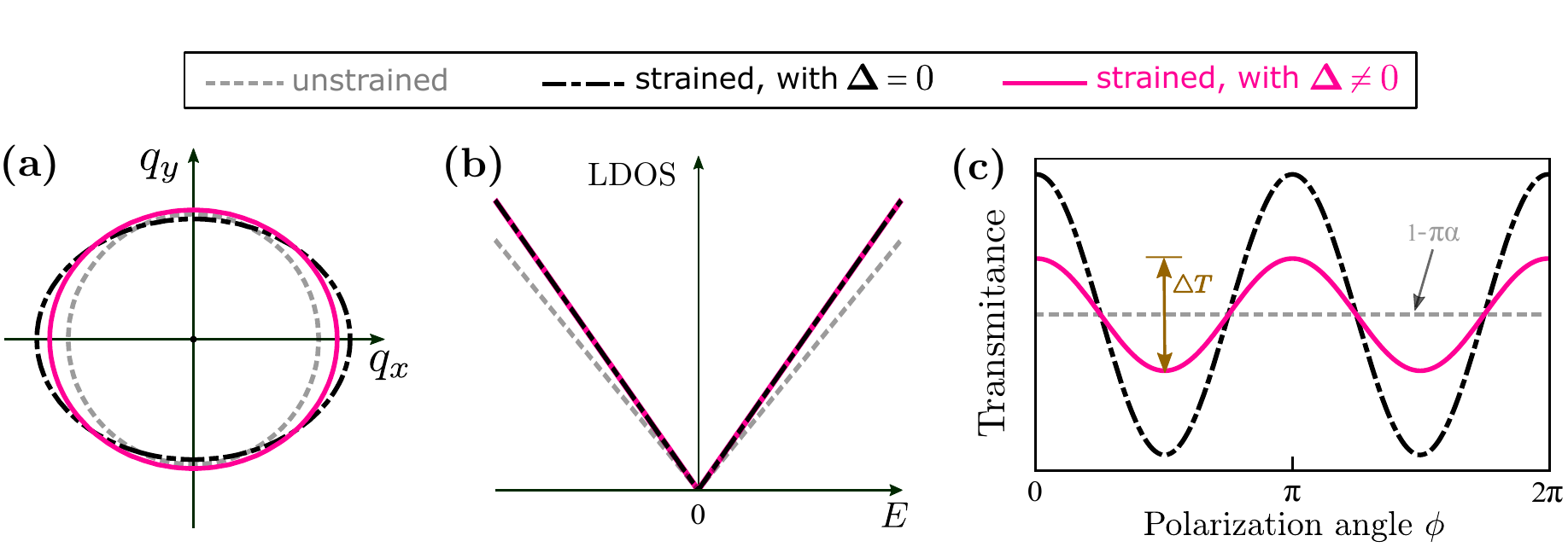}
\caption{(a) Equienergy contours around the Dirac points, (b) local densities of states (LDOS) in arbitrary units and (c) light transmittances for graphene either unstrained (gray dashed lines) or uniaxially strained, such that $\epsilon_{xx}=\varepsilon$, $\epsilon_{yy}=-\nu\varepsilon$ and $\epsilon_{xy}=0$. For each panels, the black short-long dashed lines correspond to those from the standard Cauchy-Born rule with $\bm{\Delta}=0$, while pink solid lines show results from the general transmorfation rule (\ref{GCB}). The used parameters are $\varepsilon=0.1$, $\nu=0.16$, $\beta=3$ and $\kappa=2/5$.\label{fig2}}
\end{figure*}

A simple exploration shows that equation~(\ref{OCT-G}) reproduces, for $\kappa=0$, the previous results obtained within the approximation $\bm{\Delta}=0$ \cite{Oliva2014,Oliva2017a}. Moreover, such as occur for the position of the Dirac points, the more general conductivity tensor (\ref{OCT-G}) can be obtained from the expression for the optical conductivity derived in Ref.~\cite{Oliva2014} by means of the simple replacement $\beta\rightarrow\beta(1-\kappa)$. Returning to the example of a uniaxial strain, it follows from equation (\ref{OCT-G}) that the optical conductivity perpendicular to the stretching direction $\sigma_{\perp}=\sigma_{0}(\omega)[1 + \beta^{*}\varepsilon(1+\nu)]$ increases  by the same amount that the parallel conductivity $\sigma_{\parallel}=\sigma_{0}(\omega)[1 - \beta^{*}\varepsilon(1+\nu)]$ decreases. Actually, this increase-decrease balance is broken whereas second-order terms of the strain tensor are taken into account, because the trigonal symmetry of the underlying honeycomb lattice is revealed \cite{Pereira2010,Oliva2017a}. 

An observable consequence of the anisotropic optical response of strained graphene is the periodic modulation of its transmittance as a function of the light polarization direction \cite{Ni2014}. In particular, for normal incidence of linearly polarized light on graphene in vacuum, its transmittance $T$ under uniaxial strain is given by
\begin{equation}\label{T}
T = 1 - \pi\alpha[1 - \beta^{*}(1+\nu)\varepsilon\cos2\phi],
\end{equation}
where $\alpha\approx1/137$ is the fine-structure constant and $\phi$ is the angle formed by the incident-light polarization and the stretching direction. Thus, the transmittance modulation amplitude results $\triangle T = 2\pi\alpha\beta^{*}(1+\nu)\varepsilon$, which allows to estimate the stretching magnitude $\epsilon$ from the measurement of $\triangle T$ \cite{Ni2014}. Note that such procedure would underestimate the value of $\epsilon$ if the relative displacement vector $\bm{\Delta}$ is not considered. 

Moreover, equation (\ref{T}) suggests that the effect of $\bm{\Delta}$ should be detectable by means of transmittance experiments. For example, using typical parameters $\beta\approx3$ and $\kappa\approx2/5$ \cite{Midtvedt2016}, the resulting transmittance modulation amplitude $\triangle T$ would be $40\%$ of its predicted value according to Refs. \cite{Pereira2010,Oliva2015b}, as illustrated in Fig. \ref{fig2}(c). Therefore, such type of experiment could confirm the presence of the relative displacement between sublattices $\bm{\Delta}$, if $\beta$ is previously determined, for instance, by STS measurements.

\section{Conclusions} 

In closing, we have studied the low energy electronic properties of graphene under uniform strain by assuming that the nearest-neighbor vectors transform according to the new rule $\bm{\delta}_{n}\rightarrow(\bar{\bm{I}} + \bar{\bm{\epsilon}})\cdot\bm{\delta}_{n} + \bm{\Delta}$ that goes beyond the commonly used Cauchy-Born rule \cite{Midtvedt2016}. Due to the consideration of the strain-induced relative displacement vector $\bm{\Delta}$ between the two sublattices, the new obtained effective Dirac Hamiltonian $\mathcal{H}=\hbar \bm{\sigma}\cdot\bar{\bm{v}}\cdot\bm{q}$ for strained graphene presents a Fermi velocity tensor $\bar{\bm{v}}$ given by equation~(\ref{TFV}) with a qualitatively different behavior as a function the strain tensor. For example, under uniaxial strain, the derived $\bar{\bm{v}}$ here predicts that the Fermi velocity perpendicular to the stretching direction decreases with increasing strain magnitude. 

Moreover, we have analyzed the effects of $\bm{\Delta}$ on measurable quantities of strained graphene, such as the LDOS and the optical conductivity. As discussed, from STS experiments, without and with the presence of a uniform magnetic field, one can not observe fingerprints of $\bm{\Delta}$ because, at least up to the first order in the strain tensor, the LDOS is not modified by the occurrence of such relative displacement between sublattices. This fact reveals that either standard STS or Landau level spectroscopy could be adequate techniques to determine $\beta$ that usually is estimated from ab initio calculations.

In contrast, we have demonstrated that the optical conductivity tensor (\ref{OCT-G}) does record effects of the relative displacement $\bm{\Delta}$, because it has the same functional form that previous expression obtained for $\bm{\Delta}=0$, but with renormalized parameters, i.e. $\beta$ by $\beta(1-\kappa)$. This finding allows the use of transmittance experiments to unveil the generalized Cachy-Born rule (\ref{GCB}). As a consequence, the effect of $\bm{\Delta}$ should be considered for a more complete interpretation of the optical measurements of strained graphene.
 
\begin{acknowledgments}
This work has been partially supported by CONACyT of Mexico through Project 252943, by PAPIIT of Universidad Nacional Aut\'onoma de M\'exico (UNAM) through Project IN106317, and by Miztli-LANCAD of UNAM. M.O.L. acknowledges the postdoctoral fellowship from DGAPA-UNAM.
\end{acknowledgments}

\bibliography{biblioStrainedGraphene}

\end{document}